\documentclass[epj,referee]{svjour}
\usepackage{graphics}

\usepackage{amsfonts}
\usepackage{epsfig}
\usepackage{amsmath}
\usepackage{times}

\begin{document}

\title{Uncovering Evolutionary Ages of Nodes in Complex Networks}

\author{Guimei Zhu\inst{1,2}\and Huijie Yang\inst{3,2,} \thanks{\email{hjyang@usst.edu.cn}}\and Rui Yang\inst{4}\and Jie Ren\inst{2,1}\and Baowen Li\inst{2,1,} \thanks{\email{phylibw@nus.edu.sg}}\and Ying-Cheng
Lai\inst{4,5,2,} \thanks{\email{Ying-Cheng.Lai@asu.edu}}}


\institute{NUS Graduate School for Integrative Sciences and
Engineering, Singapore 117456, Republic of Singapore \and Department
of Physics and Center for Computational Science and Engineering,
National University of Singapore,Singapore 117546 \and School of
Business, Shanghai University for Science and Technology, Shanghai
200092, China \and School of Electrical, Computer, and Energy
Engineering, Arizona State University, Tempe, AZ 85287, USA \and
Department of Physics, Arizona State University, Tempe, AZ 85287,
USA}

\date{Received: date / Revised version: date}

\abstract{
In a complex network, different groups of nodes may have
existed for different amounts of time. To detect the evolutionary
history of a network is of great importance. We present a
spectral-analysis based method to address this fundamental question
in network science. In particular, we find that there are complex
networks in the real-world for which there is a positive correlation
between the eigenvalue magnitude and node age. In situations where
the network topology is unknown but short time series measured from
nodes are available, we suggest to uncover the network topology at
the present (or any given time of interest) by using compressive
sensing and then perform the spectral analysis. Knowledge of ages of
various groups of nodes can provide significant insights into the
evolutionary process underpinning the network. It should be noted,
however, that at the present the applicability of our method is
limited to the networks for which information about the node age has
been encoded gradually in the eigen-properties through evolution.
} 

\maketitle

\section{Introduction} \label{sec:intro}

  Many large, complex networks in existence today are the results of some
evolutionary processes such as growth \cite{Newman:book}. The Internet is one
best example, which has undergone tremendous expansion in the past two decades.
Growth in a decentralized manner also appears to be the hallmark of other types
of networks such as various biological, social and
economical networks (e.g., Facebook). Given a complex network but without
any knowledge of its evolutionary history, one might be interested in
the distribution of the ``ages'' of various nodes or subgroups of nodes
in the network. Information about the node ages can provide deep
insights into the organization and structure of the underlying network,
and may have significant applications. For example, in a social network,
the lifetimes of certain subgroups of nodes may be closely related to
the network backbone structure in terms of the roles
that these subgroups play in the function of the network, e.g.,
leadership roles. In a biological network, nodes of longer lifetimes
can be more critical to the various functions of the network. It is
thus of considerable interest to develop a systematic method to uncover
the evolutionary ages of subgroups of nodes in complex networks.

  Two situations arise when addressing the age-detection problem in
complex networks: (1) network topology is known and (2) the topology
is unknown but only time series measured or observed from various
nodes are available. In the first case we shall establish that the
spectrum of the network connectivity matrix, or the Laplacian
matrix, is directly related to the evolutionary ages of various
subgroups of nodes in the network. In the second case, we make use
of a recently developed method of time-series based reverse
engineering of complex networks \cite{WYLHK:2011} to uncover the
network topology, and then could analyze the spectrum of the
predicted Laplacian matrix to obtain estimates of the age
distribution of nodes. Our approach thus defines a framework in
which the problem of evolutionary-age detection of nodes in complex
networks can be addressed in systematic way. While our method does
not require a positive correlation between the node degree and age,
a correlation between the eigenvalue and the node age is necessary.

  It is useful to point out that for the class of scale-free networks
that are generated according to the preferential-attachment rule
\cite{BA:1999}, the problem of evolutionary-age estimation may be
trivial. In particular, this growth rule stipulates that the probability
for an existing node to acquire new links is proportional to its degree,
implying a strong correlation between the node degree and its lifetime.
Thus, for a scale-free network evolved predominantly according to the
preferential-attachment rule, the ages of various nodes can be predicted
simply by examining the degrees. However, many real-world networks
deviate significantly from the scale-free topology \cite{Newman:book}
and, for them the problem of detecting node evolutionary ages is
nontrivial. Nonetheless, scale-free networks provide an ideal testbed
to validate our spectrum-analysis method.

  We emphasize that, although our method is suitable even for networks for
which there is no positive correlation between node degree and age, its
applicability is limited to networks for which there is a
positive correlation between the properties of the eigenmodes and the node
age. For networks with which no evolutionary process can be affiliated, such
as various citation networks and twitter-type of social networks where the
importance of a node may not be related with its age, our method
is not applicable.

  In Sec. \ref{sec:method}, we describe the main idea underlying our
method. In Sec. \ref{sec:SF}, we validate the method by using
scale-free networks generated by the standard
preferential-attachment rule and by the duplication/divergence
mechanism, which are especially relevant to social and biological
systems, respectively. In Sec. \ref{sec:PPI}, we consider a
realistic biological network, the protein-protein interaction
network for which the age distribution of nodes is available, to
further validate our method. In Sec. \ref{sec:CS}, we address the
situation where the network topology is not known {\em a priori} but
only time series are available, make use of the reverse-engineering
approach \cite{WYLHK:2011} to map out the network topology, and
demonstrate that the approach yields correctly and accurately the
spectrum of the Laplacian matrix. A brief conclusion is presented in
Sec. \ref{sec:conclusion}.

\section{Method} \label{sec:method}

  For a complex network of $N$ nodes, its topological structure can be
described by the Laplacian matrix $L$
\cite{ZHU:2008,YANG:2006,REN:2009,REN:2010,SARIKA:2011}, where the
off-diagonal elements of $L$ are $L_{i\ne j}=L_{j\ne i} = 1 (0)$ if
the nodes $i$ and $j$ are connected (disconnected), respectively.
The diagonal elements are $L_{ii}=-\sum_{j\ne i}L_{ij}=-k_{i}$,
where $k_i$ is the number of the nodes connected directly with the
node $i$ (node degree). The eigenvalues of $L$ are nonnegative and
can be ranked as $0 = \lambda_1 \le \lambda _2 \cdots \le \lambda
_N$. The corresponding eigenvectors are $X_1 ,X_2 , \cdots ,X_N $,
whose wavelengths are sorted in a descending order. Each eigenvector
contains components concentrated on various nodes in the network.

\begin{figure}
\begin{center}
\scalebox{0.75}[0.75]{\includegraphics{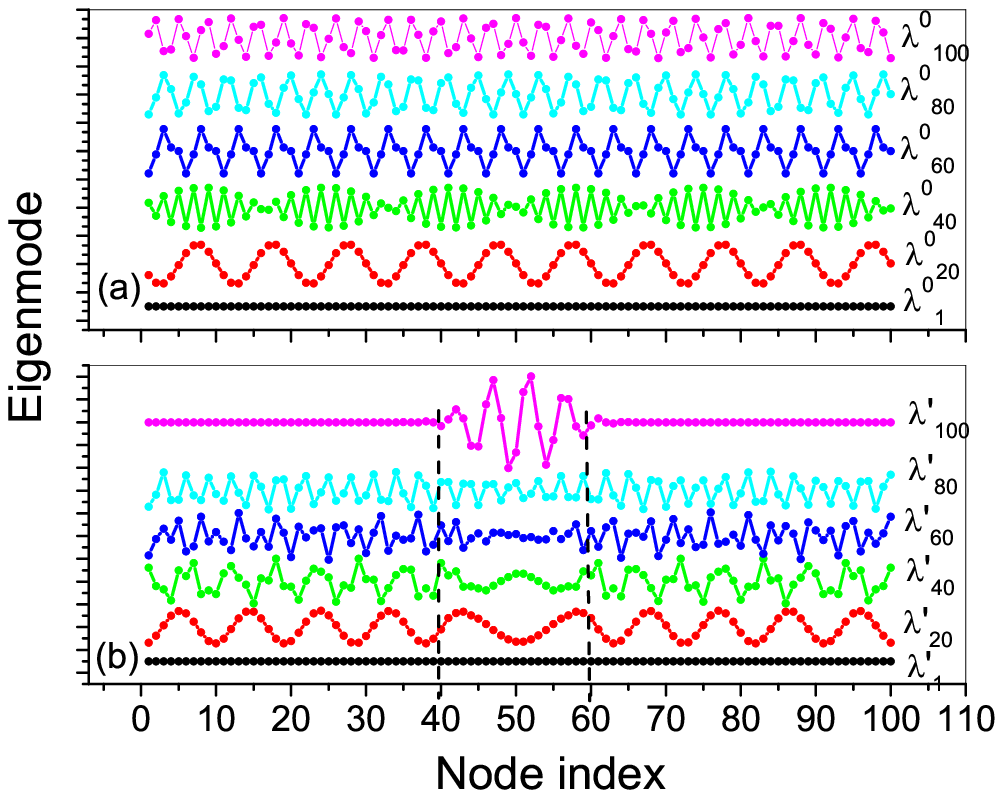}}
\scalebox{0.75}[0.75]{\includegraphics{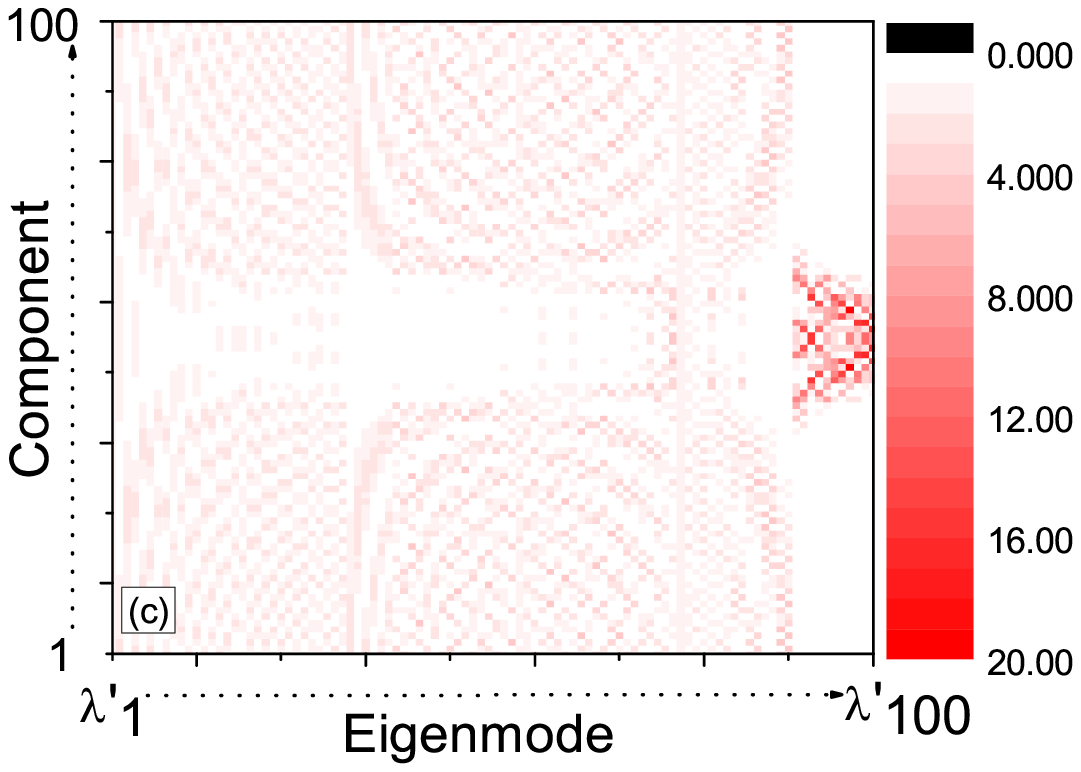}}
\end{center}
\caption{(color online.) For a regular ring network of 100 nodes
where each node has four neighbors, (a) examples of typical,
periodic-wave like eigenvectors, (b) typical eigenvectors when
each node in the group of indices between 40 and 60 acquires
two additional links, one on each side. We observe significant
distortions from the periodic-wave pattern, which are localized
between the 40th and 60th components of eigenvectors associated
with relatively large eigenvalues. (c) Representation of all
eigenvectors, where those associated with eigenvalues from $\lambda'_{90}$
to $\lambda'_{100}$ are significantly more sensitive to the structural
perturbation to the network.}
\label{fig:lattice}
\end{figure}

  For a regular or a small-world
network \cite{WS:1998}, the eigenvectors typically exhibit some
wave patterns with certain wavelengths \cite{ROH:2004,PHL:2007}.
When a perturbation is applied to the network, the affected
eigenvectors are those whose wavelengths match the size of the
perturbation (i.e., the number of nodes that it affects). In this case,
some localized structure in the affected eigenvectors can emerge.
Eigenvectors associated with small eigenvalues usually have large
wavelengths, and so they are sensitive to perturbation on a global
scale. In contrast, eigenvectors associated with large eigenvalues
are most sensitive to localized perturbations that are applied to a
small set of nodes in the network. The responses of the eigenvectors
to perturbations thus reflect the structure of the network at different
scales. An example is given in Fig. \ref{fig:lattice} for a
one-dimensional regular lattice of $N = 100$ nodes with periodic boundary condition,
where each node is connected with 2 neighbors on either side so that
the node has 4 nearest neighbors. Shown in Fig. \ref{fig:lattice}(a)
are representative eigenvectors, where the values of $N \cdot X_i^2(s)$
are plotted and $X_i^2(s)$ is the $s$th component of the eigenvector
$X_i$. We see that the eigenvectors represent periodic waves of
wavelengths ranging from $N$ to $2$. To observe
the effect of local structural perturbation on the eigenvectors, we
add two more links to each node in the group of nodes whose indices
are between 40 and 60 so that each node in this perturbed group now
has six nearest neighbors. Let $\lambda'_i$ ($i = 1, \ldots, N$) be the
eigenvalues in the perturbed network. Figure \ref{fig:lattice}(b)
shows some representative eigenvectors. We observe that the eigenvectors
associated with small eigenvalues, e.g., $\lambda'_1$, $\lambda'_{20}$,
$\lambda'_{40}$, $\lambda'_{60}$, and $\lambda'_{80}$, are basically
unchanged. However, eigenvectors associated with relatively large
eigenvalues, such as $\lambda'_{100}$, are strongly altered
by the perturbation but the changes are focused on the perturbed group
of nodes. Figure \ref{fig:lattice}(c) shows the distribution of
the magnitudes of all eigenvectors on nodes in the network, where
we see that those associated with eigenvalues $\lambda'_{90} $ to
$\lambda'_{100}$ are sensitive to the perturbation with large variations
appearing on the perturbed nodes.

  For complex networks that do not possess a regular backbone, such as
random \cite{ER:1960} and scale-free \cite{BA:1999} networks, the
eigenvectors in general do not exhibit any periodic wave structure.
Nonetheless, the observation that the eigenvectors associated with larger
eigenvalues are more sensitive to structural perturbations can be
used to infer the evolutionary age of nodes. To see this, consider
a scale-free network evolved according to the preferential-attachment
rule \cite{BA:1999}, for which there is a positive correlation between
the node degree and lifetime. That is, nodes of ``old'' ages tend to
have more links and they are thus more susceptible to perturbations
applied randomly to the network during the evolutionary process.
Since the eigenvectors of large eigenvalues are quite sensitive to
perturbations (c.f., Fig. \ref{fig:lattice}), we expect the large-degree
nodes to dominate these eigenvectors. As a result, large eigenvalues
tend to correspond to nodes of long lifetime. This argument suggests
that, nodes having the most significant components of the eigenvectors
associated with the largest eigenvalues are likely to possess the longest
lifetime in the network.

\section{Validation using scale-free networks} \label{sec:SF}

\begin{figure}
\begin{center}
\scalebox{0.4}[0.4]{\includegraphics{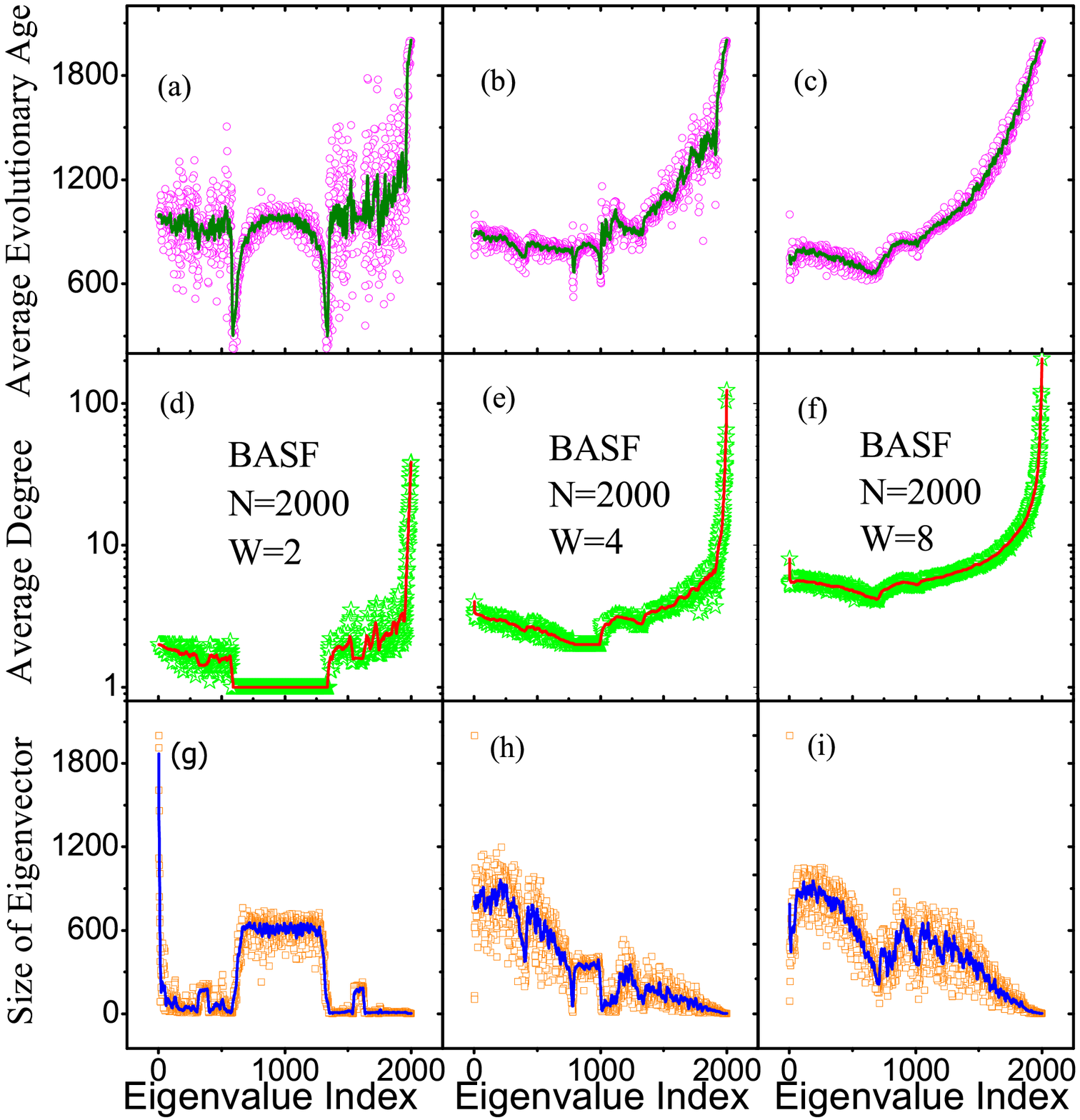}}
\scalebox{0.3}[0.3]{\includegraphics{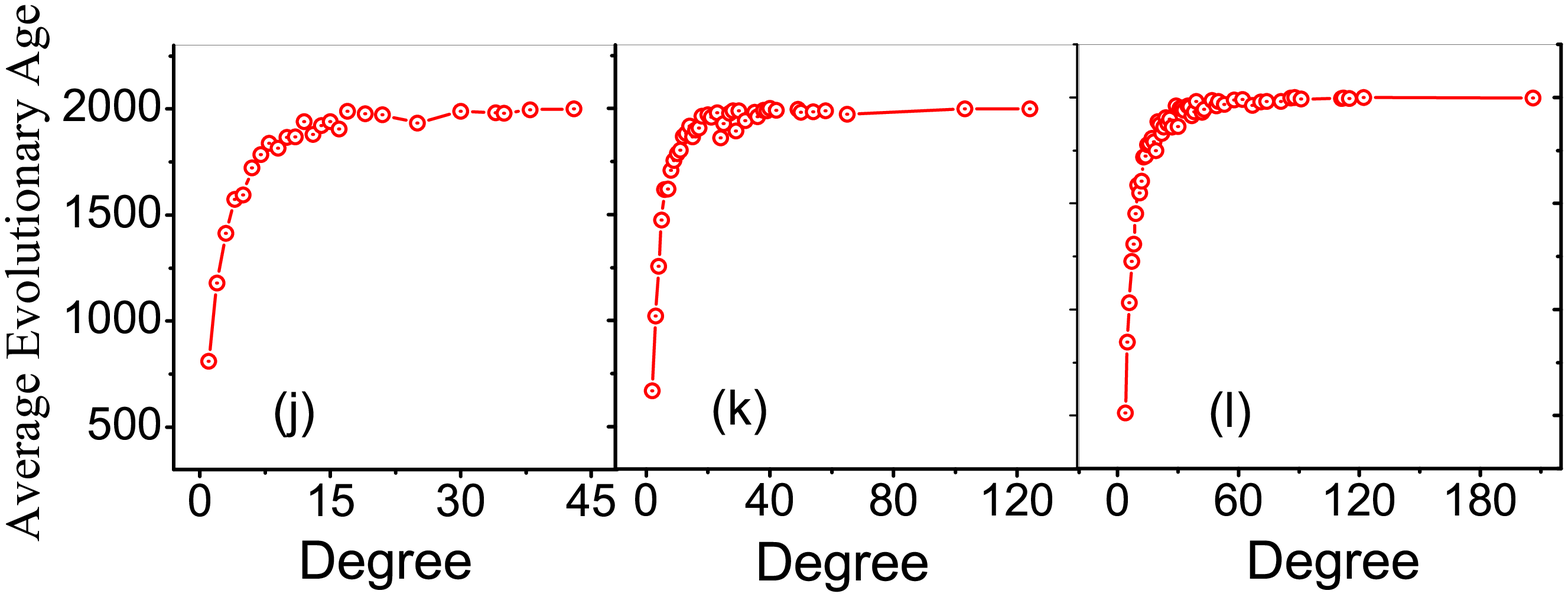}}
\end{center}
\caption{(Color online.) For three scale-free networks generated according to
the standard preferential-attachment rule with edge density $w = 2,4,8$
(corresponding to the left, middle, and right column, respectively),
(a-c) average ages, (d-f) average degree (on a logarithmic scale), and (g-i) size of
eigenvector versus the eigenvalue index $i$. Eigenvectors associated with
large eigenvalues generally have small sizes, but their ages are ``older''
in the network. (j-l) Average age versus degree. We see that, while small
degree is related with the average age, information about node age deteriorates
quickly as the degree is increased.}
\label{fig:SFN}
\end{figure}

  To exemplify the relation between eigenvalues and node ages, we
consider standard scale-free networks \cite{BA:1999}. Each network
has $N = 2000$ nodes, which is evolved following the
preferential-attachment rule so that the age of the $i$th node is $N-i+1$.
For a given eigenvalue, the lifetime of the associated eigenvector
is the average age of all nodes contained in the vector, weighted by
the respective components of the eigenvector. Figures \ref{fig:SFN}(a-c)
show the ages of the eigenvectors $X_i$ versus the index $i$ for three
networks of different edge density $w$. The significant feature common
to all three cases is that the average age of the nodes
dominating some eigenvector increases on
average with the eigenvalue. The average degree of each eigenvector,
i.e., the weighted average of the degrees of all nodes associated with
the vector, shows the same tendency, as shown in Figs. \ref{fig:SFN}(d-f),
where the average degree is presented on a logarithmic scale. For each
network, the sizes of the eigenvectors are shown in Figs. \ref{fig:SFN}(g-i),
where the size of an eigenvector is defined to be the number of nodes
on which the vector component is larger than a small threshold value.
For sufficiently dense network, e.g., Fig. \ref{fig:SFN}(i), the size
tends to decrease on average with the eigenvalue, indicating
that a small group of nodes have extraordinarily long lifetimes
in the network and their relative ages can be identified simply by
examining the associated eigenvalues. Figures \ref{fig:SFN}(j-l) show,
for $W=2,4$ and $8$, respectively, the average evolution age versus
the node degree. We observe an approximately monotonic relation for
small degree. However, when the node degree is larger than 10, the
relation deteriorates quickly and the relations approach a constant.

\begin{figure}[tbp]
\begin{center}
\epsfig{figure=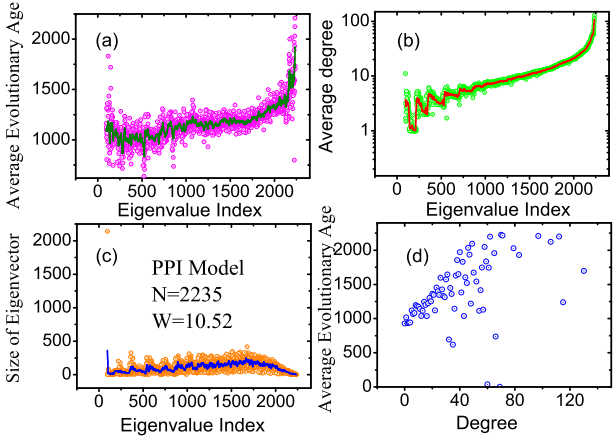,width=\linewidth}
\end{center}
\caption{(Color online.) For scale-free networks generated by
duplication/divergence-based mechanism from PPI network of the
Baker's Yeast, (a) average age versus eigenvalue index,
(b) average degree versus eigenvalue index, and (c) size of
eigenvector versus the eigenvalue index. Eigenvectors associated
with large eigenvalues generally have small sizes, but their ages
are ``older'' in the network. (d) Average age versus degree. Because
of large fluctuation, the degree cannot give age-related
information, but the eigenvalues can.} \label{fig:PPIB}
\end{figure}

  To further demonstrate our method, we have analyzed a scale-free
cellular network generated by mechanism different than that of
the preferential-attachment rule, namely the
protein-protein interaction(PPI) networks. In such a network, duplication and
divergence are believed to be responsible for the topological structure
\cite{PPIB:2003}. We start from a small, connected graph as a seed and duplicate
a randomly selected existing protein at each step. The new comer duplicates exactly
the connection pattern of its generator in the network. Due to mutations,
some of the duplicated edges are broken with probability $p$, while new edges
are generated with probability $q$ between the new comer and other existing nodes.
To compare with the PPI network of the Baker's Yeast (to be described in
the next Section), we generate networks with comparable parameters. In particular,
a typical network has $2235$ nodes and average degree of $10.52$, and degree
distribution follows power-law with exponent $2.3$. In a wide range of eigenvalues
there exists a strong correlation between the eigenvalue and average age, as shown
in Fig.\ref{fig:PPIB}(a). We observe that, the curve of average age versus degree exhibits
large fluctuations, as shown in Fig.\ref{fig:PPIB}(d). It is thus not possible
to obtain information about node age from degree. However, behaviors of
the eigenmodes can reveal the age information, as will be demonstrated
in Sec. \ref{sec:PPI}.

\section{Evolution ages of nodes in a protein-protein interaction network}
\label{sec:PPI}

\begin{figure}[tbp]
\begin{center}
\epsfig{figure=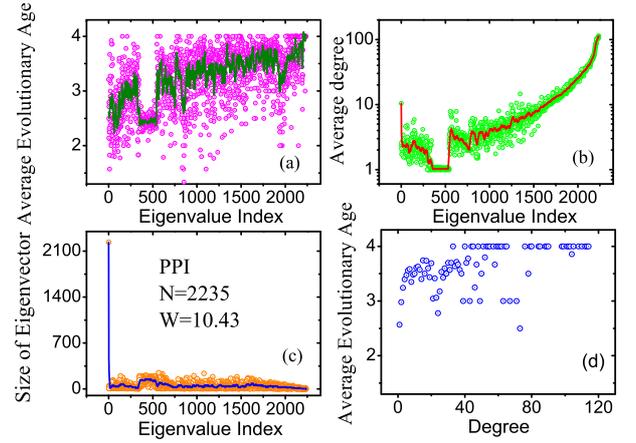,width=\linewidth}
\end{center}
\caption{(Color online.) For the largest connected component of the
PPI network of the baker's yeast with 2235 nodes, (a) the
evolutionary age, (b) average degree (on a logarithmic scale), and
(c) size of eigenvector versus the eigenvalue index $i$. These
results further indicate that the evolutionary ages of various nodes
in the network can be inferred from the eigenvalue spectrum of the
Laplacian matrix. (d) Average age versus degree. We see that degree does
not reveal age-related information.} \label{fig:PPI}
\end{figure}

  To lend more credence to our proposition that the evolutionary ages of
nodes can be inferred from the eigenvalues, we now consider a class
of networks in systems biology, protein-protein interaction (PPI)
networks. These networks are the result of a number of evolutionary
mechanisms such as duplications of genes and reattachments of links
between the proteins. Specifically, we analyze the PPI network of
the baker's yeast (\textit{Saccharomyces cerevisiae})
\cite{wangner01,wangner03}. Von Mering et al. \cite{mering02}
analyzed a total of 80000 interactions among 5400 yeast proteins
reported previously and assigned each interaction a confidence
value. In order to reduce the effect of false positives, we focus on
11855 interactions with high and medium confidence values among 2617
yeast proteins. In a PPI network, each protein is a node and each
pairwise interaction represents a link between two nodes. Since our
goal is to assess, through the eigenvalues, the evolutionary ages of
the nodes, we neglect the directions of the edges. The largest
connected component of the PPI network contains 2235 nodes. In
systems biology, the evolutionary processes of the proteins are
classified into four iso-temporal groups \cite{woese87}:
prokaryotes, eukarya, fungi, and yeast, to which numbers $4,3,2$ and
$1$ are assigned according to their evolutionary process from
ancient to modern times, respectively. The evolutionary age of a
protein is the largest number from the groups it presents. For
example, the protein YHR$037$w occurs in the groups
prokaryotes($4$), eukarya($3$), fungi($2$), which means that it can
be found from the ancient prokaryotes, so that its age is $4$.
Figure \ref{fig:PPI}(a) shows the average evolutionary age of nodes
in eigenvector versus the eigenvalue index, which is similar to the
behavior in Figs. \ref{fig:SFN}(a-c). This suggests that for a
realistic biological network, there is indeed a positive correlation
between the eigenvalues of the Laplacian matrix and the evolutionary
ages of groups of nodes. Since PPIs typically possess a scale-free
structure \cite{RSMOB:2002}, we expect the average degree of groups
of nodes to exhibit similar behaviors as in Figs.
\ref{fig:SFN}(d-f). This is indeed the case, as shown in Fig.
\ref{fig:PPI}(b). The sizes of various eigenvectors are shown in
Fig. \ref{fig:PPI}(c). Again the behavior is similar to those in
Figs. \ref{fig:SFN}(g-i). From Fig. \ref{fig:PPI}(d), relation of
average age versus degree, we see that the degree contains no
information about the node age.

\section{Time-series based detection of evolutionary ages of nodes}
\label{sec:CS}

We now address the situation where the network topology is unknown
but only time series measured or observed from various nodes are
available. We shall apply a recently developed approach
\cite{WYLHK:2011} based on compressive sensing
\cite{CRT:2006,C:2006,D:2006,B:2007,C:2008,CR:2005} to uncover the
complex-network topology and then could analyze the spectrum of the
predicted Laplacian matrix to estimate the evolutionary ages of
nodes. The unique feature of compressive sensing lies in its
extremely low data requirement: very little observation is needed to
obtain a target sparse signal. In general, the problem of
compressive sensing can be described as to reconstruct a sparse
vector $\mathbf{a}\in R^N$ from linear measurements $\mathbf{X}$
about $\mathbf{a}$ in the form: $\mathbf{X}=\mathbf{G} \cdot
\mathbf{a}$, where $\mathbf{X}\in R^M$ and $\mathbf{G}$ is an
$M\times N$ matrix. Accurate reconstruction can be achieved by
solving the following convex optimization problem \cite{CRT:2006}
\begin{equation} \label{eq:1_2}
\min \|\mathbf{a}\|_1 \quad \mbox{subject \ to} \quad
\mathbf{G}\cdot \mathbf{a}=\mathbf{X},
\end{equation}
where $\|\mathbf{a}\|_1=\sum_{i=1}^{N}|\mathbf{a}_i|$ is the $L_1$
norm of vector $\mathbf{a}$ and $M\ll N$, $i.e.,$, the number of
measurements can be much less than the number of components of the
unknown signal. Various solutions of the convex optimization problem
(\ref{eq:1_2}) have been worked out in the applied-mathematics literature
\cite{CRT:2006,C:2006,D:2006,B:2007,C:2008,CR:2005}.

To uncover network topology based on data, it is necessary to cast
the problem in the form (\ref{eq:1_2}). The basic hypothesis is that
a complex networked system can be viewed as a large dynamical system
that generates oscillatory time series at various nodes. Under this
hypothesis, it is straightforward to formulate the problem under the
compressive-sensing paradigm, details of which can be found in
Ref. \cite{WYLHK:2011}.

\begin{figure}
\begin{center}
\epsfig{figure=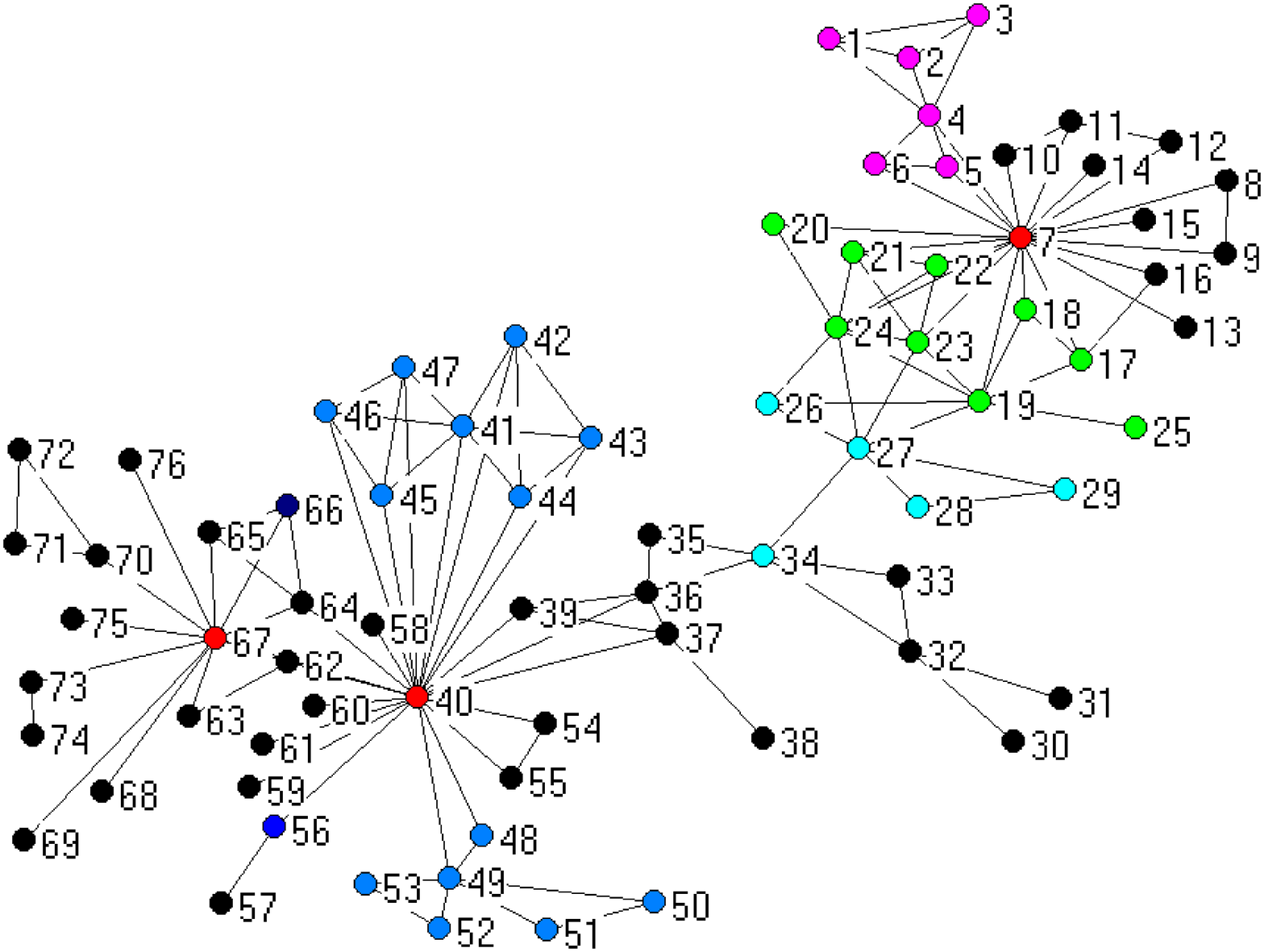,width=\linewidth}
\end{center}
\caption{(Color online.) Schematic illustration of the largest
component of the SFI collaboration network and the clustered
structure revealed by an eigenvalue/eigenvector analysis.}
\label{fig:SFIN}
\end{figure}

To give a concrete example, we consider a real-world network, the
Santa Fe Institute (SFI) collaboration network \cite{girvan02}.
There are $N =76$ nodes in the largest connected component of the
network and the average degree is about 3. A schematic illustration
of the network is shown in Fig. \ref{fig:SFIN}. A spectral analysis
reveals that the eigenvectors associated with $\lambda_{76}$,
$\lambda_{75}$ and $\lambda _{74}$ characterize the three hubs: 40,
7 and 67, all marked by red. The eigenvector associated with
$\lambda_{73}$ involves a group of nodes numbered between 17 and 25
(marked by green). For $\lambda _{72}$, the corresponding
eigenvector covers nodes 26 to 29, and node 34 (marked by cyan). The
three clusters: nodes 41 to 47 (blue), 1 to 6 (magenta), and 48 to
53 (violet), are represented by eigenvectors $\lambda _{70}$,
$\lambda _{69}$, and $\lambda _{68}$, respectively. In fact,
clusters of larger scales can be identified for smaller eigenvalues.

\begin{figure}[tbp]
\begin{center}
\epsfig{figure=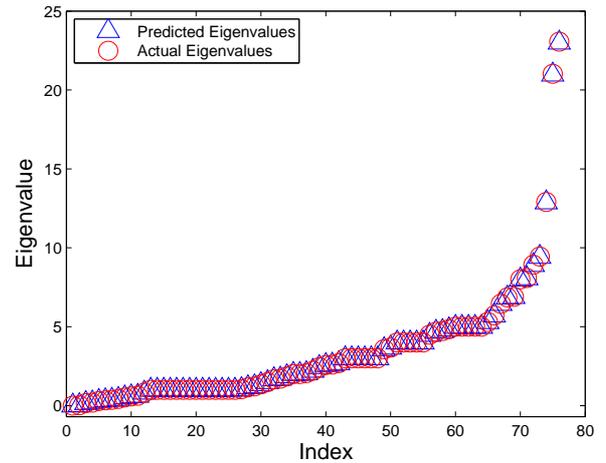,width=\linewidth}
\end{center}
\caption{(Color online) Sorted eigenvalues of the predicted and
actual Laplacian matrix of the SFI collaboration network. The number
of data points used in uncovering the network structure is about
$40\%$ of the number of total unknown coefficients in the
power-series expansion.} \label{fig:CS_eigen}
\end{figure}

Now assume that the network topology is unknown but an oscillatory
time series from each node is available. To simulate the situation,
we assume that the dynamics of each node is described by the chaotic
R\"{o}ssler oscillator \cite{Rossler:1976}. Applying the
compressive-sensing based method to uncover the network topology,
we can then perform a spectral analysis to estimate the ages of
various nodes in the network. Figure \ref{fig:CS_eigen} shows the
eigenvalues of the predicted and the actual Laplacian matrix. We
observe an excellent agreement.

\section{Conclusions} \label{sec:conclusion}

In summary, we have developed a procedure to estimate the
evolutionary ages of nodes in complex networks. The basic
observation is that eigenvectors associated with different
eigenvalues of the Laplacian matrix can typically represent highly
localized groups of nodes in the network. A qualitative argument can
then be made for the existence of positive correlation between the
node ages and the magnitudes of the eigenvalues. This means that,
when the network topology is known, a simple eigenvalue analysis can
lead to reliable information about the age distribution of nodes in
the network. For situations where the network topology is unknown
but time series from nodes are available, it is necessary to uncover
the topology in order to estimate the node ages, and we have
demonstrated that this can be done efficiently using compressive
sensing. Examples from model and real-world networks, including a
PPI network, are used to validate our approach. We hope our method
to find applications in fields such as systems biology,the
propagation of a rumor, a fashion, a joke, or a flu, where
estimating node ages can be of significant value.

The network-reconstruction technique used in our work is based on compressive sensing,
which works for situations where the types of mathematical forms
of the nodal dynamical systems and coupling functions are known
(although details of these functions are not required) and can
be represented by series expansion. So far the method has not
been applied to gene-regulatory networks due to difficulty to
find suitable series expansions. The recent method by Hempel
et al. \cite{HKKN:2011} is based on extracting statistical information and has
been demonstrated to work well for gene-regulatory networks.

While many real-world systems such as gene regulatory and supply chain
networks are directed, our present work focused on undirected networks.
The main consideration is that many networks generated by some kind of
evolutionary processes or constructed through experiments tend to
undirected. For example, the Baker Yeast obtained through the approach
of prey and predator contains no information about the directionality
of the nodal interactions. Our method is based on the observation that
local structures, e.g., densely connected clusters, can induce large
components in the eigenvectors. Hubs or clusters of hubs can then be
detected by the eigenvectors corresponding to largest eigenvalues,
while clusters of larger sizes can be uncovered by eigenvectors of
smaller eigenvalues. Different eigenmodes can be used to detect
clusters of varying scales, providing a correlation with the
evolutionary ages in situations where hubs or clusters of hubs
are formed by history. The principle on which our method is based
thus does not take into account directionality in the node-to-node
interactions. To develop a method to uncover the evolutionary ages
for directed complex networks remains to be an interesting but open
question at the present.

\section*{Acknowledgement}

HJY was supported by the National Science Foundation of China under
Grants No. 10975099 and 10635040, the Program for Professor of
Special Appointment (Eastern Scholar) at Shanghai Institutions of
Higher Learning, and the Shanghai leading discipline project under
grant No.S30501. YCL was supported by AFOSR under
Grant No. FA9550-10-1-0083. We also acknowledge Mr. Liu Sha and Mr. Tao Lin for
useful discussions.


\begin{thebibliography}{}

\bibitem{Newman:book}
M. J. Newman, {\it Networks: An Introduction} (Oxford University
Press, New York, 2010).

\bibitem{WYLHK:2011}
W.-X. Wang, R. Yang, Y.-C. Lai, V. Kovanis, and M. A. F. Harrison,
Europhys. Lett. {\bf 94}, 48006 (2011); W.-X. Wang, R. Yang, Y.-C.
Lai, V. Kovanis, and C. Grebogi, Phys. Rev. Lett. {\bf 106}, 154101
(2011).

\bibitem{BA:1999}
A.-L. Barab\'asi, and R. Albert, Science {\bf 286}, 509 (1999).

\bibitem{ZHU:2008}
G. M. Zhu, H. J. Yang, C. Y. Yin, B. Li, Phys. Rev. E
\textbf{77}, 066113 (2008).

\bibitem{YANG:2006}
H. J. Yang, F. C. Zhao, and B. H. Wang, Chaos \textbf{16} (2006).

\bibitem{REN:2009}
J. Ren, and B. Li, Phys. Rev. E \textbf{79}, 051922 (2009).

\bibitem{REN:2010}
J. Ren, W. W. Wang, B. Li, and Y. C. Lai, Phys. Rev. Lett. \textbf{104}, 058701 (2010).

\bibitem{SARIKA:2011}
S. Jalan, G. M. Zhu, and B. Li, Phys. Rev. E \textbf{84}, 046107
(2011).

\bibitem{WS:1998}
D. J. Watts and S. H. Strogatz, Nature (London) {\bf 393}, 440 (1998).

\bibitem{ROH:2004}
J. G. Restrepo, E. Ott, and B. R. Hunt, Phys. Rev. Lett. {\bf 93},
114101 (2004).

\bibitem{PHL:2007}
K. Park, L. Huang, and Y.-C. Lai, Phys. Rev. E {\bf 75}, 026211 (2007).

\bibitem{ER:1960}
P. Erd\"{o}s and A. R\'{e}nyi, Publ. Math. Inst. Hung. Acad. Sci. {\bf 5}, 17 (1960).

\bibitem{PPIB:2003}
 A. V. Vazquez, A. Flammini, A. Maritan, A. Vespignani, Complexus {\bf 1}, 38 (2003).

\bibitem{wangner01}
A. Wagner, Mol. Biol. Evol. \textbf{18}, 1283 (2001).

\bibitem{wangner03}
A. Wagner, Proc. R. Soc. Lond. Ser. B-Biol. Sci. \textbf{270}, 457 (2003).

\bibitem{mering02}
C. von Mering, R. Krause, B. Snel, M. Cornell, S. G. Oliver, S. Fields,
and P. Bork, Nature \textbf{417}, 399 (2002).

\bibitem{woese87}
C. R. Woese, Microbiol. Rev. \textbf{51}, 221 (1987).

\bibitem{RSMOB:2002}
E. Ravasz, A. L. Somera, D. A. Mongru, Z. Oltvai, and A.-L. Barab\'{a}si,
Science {\bf 297}, 1551 (2002).

\bibitem{CRT:2006}
E. Cand\`{e}s, J. Romberg, and T. Tao, IEEE Trans. Inf. Theory {\bf
52}, 489 (2006); Commun. Pure Appl. Math. {\bf 59}, 1207 (2006).

\bibitem{C:2006}
E. Cand\`{e}s, in {\it Proceedings of the International Congress of
Mathematicians}, Madrid, Spain, 2006.

\bibitem{D:2006} D. Donoho,
IEEE Trans. Inf. Theory {\bf 52}, 1289 (2006).

\bibitem{B:2007}
R.G. Baraniuk, IEEE Signal Processing Mag. {\bf 24}, 118 (2007).

\bibitem{C:2008}
E. Cand\`{e}s and M. Wakin, IEEE Signal Processing Mag. {\bf 25}, 21
(2008).

\bibitem{CR:2005}
E. Cand\`{e}s, and J. Romberg, http://www. acm. caltech.edu/l1magic,
2005.

\bibitem{girvan02}
M. Girvan, and M. E. J. Newman, Proc. Natl. Acad. Sci. U. S. A.
\textbf{99}, 7821 (2002).

\bibitem{Rossler:1976}
O. E. R\"{o}ssler, Phys. Lett. A {\bf 57}, 397 (1976).

\bibitem{HKKN:2011}
S. Hempel, A. Koseska, J. Kurths, and Z. Nikoloski,
Phys. Rev. Lett. {\bf 107}, 054101 (2011).

\end{thebibliography}
\end{document}